\documentclass[11pt,a4paper]{article}
\usepackage{latexsym}
\usepackage{amssymb}
\usepackage{amsmath}
\usepackage{amsfonts}
\usepackage{mathrsfs}
\usepackage{cite}
\usepackage{graphicx}

\newcommand{\be}{\begin{equation}}
\newcommand{\ee}{\end{equation}}

\newcommand{\lb}{\label}

\def\tr{{\rm tr}}
\def\tr{{\rm tr}\,}
\def\Tr{{\rm Tr}\,}
\def\cN{{\cal N}}

\def\cA{{\cal A}}
\def\cF{{\cal F}}
\def\bea{\begin{eqnarray}}
\def\eea{\end{eqnarray}}
\def\nn{\nonumber}
\def\f{\frac}

\def\tr{{\rm tr}\,}

\def\nn{\nonumber}

\def\d{\delta}

\def\cA{{\cal A}}
\def\cB{{\cal B}}

\def\eq{\eqref}

\def\nb{\nabla}
\def\vp{\varphi}

\numberwithin{equation}{section}

\date{\it  }
\begin{document}
    \begin{titlepage}

 \begin{center}
 \vspace{1cm}
 {\Large\bf On the component structure of one-loop effective actions in $6D$, $ \cN=(1,0)$ and $\cN=(1,1)$
 supersymmetric gauge theories}

\vspace{1.5cm}
            {\bf
                I.L. Buchbinder\footnote{joseph@tspu.edu.ru }$^{\,a,b}$,
                A.S. Budekhina\footnote{budekhina@tspu.edu.ru}$^{\,a}$
                B.S. Merzlikin\footnote{merzlikin@tspu.edu.ru}$^{\,a,c}$,
                            }

\vspace{0.4cm}

            {\it $^a$ Department of Theoretical Physics, Tomsk State Pedagogical
                University,\\ 634061, Tomsk,  Russia \\ \vskip 0.15cm
                $^b$ National Research Tomsk State University, 634050, Tomsk, Russia \\ \vskip 0.1cm
                $^c$  Tomsk State University of Control Systems and Radioelectronics,\\
                634050, Tomsk, Russia\\
                            }
\end{center}
\vspace{0.4cm}

\begin{abstract}
We study the six-dimensional $\cN=(1,0)$ and $\cN=(1,1)$
supersymmetric Yang-Mills (SYM) theories in the component
formulation. The one-loop divergencies of effective action are
calculated. The leading one-loop low-energy contributions to bosonic
sector of effective action are found. It is  explicitly demonstrated
that the contribution to effective potential for the constant
background scalar fields are absent in the $\cN=(1,1)$ SYM theory.
\end{abstract}
\end{titlepage}

\setcounter{footnote}{0} \setcounter{page}{1}

\section{Introduction}

The various six-dimensional supersymmetric gauge theories attract a
certain interest in context of string/brane dynamics. It is known
that the superconformal $\cN=(2,0)$ theory of self-dual tensor
multiplet is closely related with low-energy dynamics of M5-branes
(see, e.g., for a review \cite{L,BL}). The study of the maximally
extended $\cN=(1,1)$ SYM theory in six-dimensions is motivated by
the connection with D5-branes (see \cite{GK} for a review). Both the
$\cN=(2,0)$ superconformal field theory and the $\cN=(1,1)$ SYM
theory were also considered as low energy limits for the little
string theories with corresponding $\cN=(2,0)$ and $\cN=(1,1)$
supersymmetries (see \cite{Ah,K}).

The $\cN=(1,1)$ SYM theory is a maximally extended supersymmetric
gauge theory of vector multiplet in six dimensions. This theory can
be treated as the $\cN=(1,0)$ vector multiplet theory coupled to
hypemultiplet which transforms under the adjoint representation of
gauge group. Although the $\cN=(1,1)$ theory is non-renormalizable
by power counting, it possesses the remarkable properties in quantum
domain and it is the subject of comprehensive research. It was
proved that the theory under consideration is on-shell finite at
one- and two loops \cite{FT,MS1,MS2,HS,HS1,BHS,BHS1} and the first
divergences here can appear only starting from three loops (see e.g.
\cite{Bork} and the references therein). Moreover it was shown,
using the $6D$ harmonic superspace approach
\cite{BIKOS,GIOS,HST,HSW,Zupnik,ISZ,IS,BIS}, that the one- and some
two-loop divergent contributions to effective action in $\cN=(1,1)$
SYM theory are finite even off-shell \cite{BIMS2,BIMS3,BIMS4} in
Fermi-Feynman gauge\footnote{The study of the gauge dependence of
the one-loop divergent contributions to one-loop effective actions
in $\cN=(1,0)$ and $\cN=(1,1)$ SYM was recently done in
\cite{BIMS5,BIMS6,BIMS7}}. The finite leading low energy
contributions to one-loop effective action were recently calculated
in superfield approach as well \cite{BIM}. Also we point out that
the essential progress in studying of the $\cN=(1,1)$ theory was
achieved in analysis of four-point on-shell scattering amplitudes
\cite{Yin1,Yin2,Amplitudes1,Amplitudes2,Amplitudes3}. In particular,
the leading and subleading divergences in all loops are obtained in
the framework of the spinor-helicity and on-shell supersymmetric
formalism (the recent results are presented in the review
\cite{Amplitudes4}).

In the present paper we study the quantum aspects of the $\cN=(1,0)$
and $\cN=(1,1)$ gauge theories in the framework of component
approach. As well known, in supersymmetric quantum field theory the
component and superfield approaches complement each other. The
component formulation of SUSY theories, although does not possess
the manifest supersymmetry, very closely relates with conventional
field theory and allows to analyze and test independently the
results which where obtained in superfield approach. Besides, the
component approach allows to use efficiently in SUSY theories the
different special methods well developed in conventional quantum
field theory. In four dimensional SUSY theories the component and
superfield approaches were developed in parallel. As to
six-dimensional SUSY theories, the component analysis of off-shell
quantum effective action is not well worked out in many details.

In this paper we are going to focus on the component calculations of
the one-loop divergences in $\cN=(1,0)$ and $\cN=(1,1)$ gauge
theories and derive the leading low-energy contribution to one-loop
effective action in $\cN=(1,1)$ theory in bosonic sector. Also we
will demonstrate the absence of the one-loop contribution to
effective potential for constant background scalar fields in the
$\cN=(1,1)$ SYM theory.

The  paper is organized as follows. In section 2 we  briefly discuss
some details of the supersymmetry in six dimensions. Also we present
an independent derivation of the action for six-dimensional ${\cal
N}=(1,0)$ SYM theory in terms of physical component fields and the
corresponding supersymmetry transformations of these fields. In
section 3, using the background field method, we evaluate the
one-loop effective action. Then we calculate the logarithmic
divergent contributions to one-loop effective action in the
$\cN=(1,0)$ and $\cN=(1,1)$ SYM theories. Section 4 includes the
details of calculation of leading low-energy contributions to
one-loop effective action and one-loop contribution to the effective
potential. The last section contains discussion of obtained results
and some possible further directions of the work.

\section{Six-dimensional $\cN=(1,0)$ SYM theory}

We begin with discussing the classical six-dimensional $\cN=(1,0)$
SYM theory. The $\cN=(1,0)$ SYM theory in six dimensions describes
the interaction of vector multiplet with a set of hypermultiplets.
The six-dimensional on-shell vector multiplet consists of real
vector field  $A_M$ and a left-handed $\lambda^{a}_i$ pseudoreal
Weyl spinor one, both in adjoint representation of gauge group. The
on-shell hypermultiplet contains complex scalar $\vp^{{\cA} i}$ and
right-handed $\psi^{\cA}_{a}$ pseudoreal Weyl spinor fields. We
denote the indices of spinor representation by the small latin
letters $a,b,..=1,2,3,4$ and for Minkowski space-time indices stand
the capital ones $M,N,..=0,1,..,5$. The indices $i,j,..=1,2$
correspond to $SU(2)$ group of $R$-symmetry of $\cN=(1,0)$
superalgebra in six dimensions. By the calligraphic letters ${\cA},
{\cB}, ..$ we denote the Pailu-G\"ursey $SU(2)$ indexes. Both
$i,j,..$ and ${\cA}, {\cB},..$ are lowered and raised by
antisymmetric quantities $\epsilon_{ij}$ and $\epsilon_{\cA\cB}$
correspondingly.

We use the antisymmetric representation of six-dimensional Weyl
matrices
 \bea
 (\gamma_M)_{ab} = - (\gamma_M)_{ba}\,, \quad
 (\tilde{\gamma}_M)^{ab} = \tfrac12\varepsilon^{abcd}(\gamma_M)_{cd}\,,
 \eea
where $\varepsilon^{abcd}$ is the totally antisymmetric tensor. The
matrices $\gamma_M$ and $\tilde{\gamma}_M$ subject to basic
relations for Weyl matrices
 \bea
 (\gamma_M)_{ac} (\tilde{\gamma}_N)^{cb}+ (\gamma_N)_{ac} (\tilde{\gamma}_M)^{cb} = -2 \delta_a{}^b \eta_{MN},
 \qquad (\gamma^M)_{ac} ({\gamma}_M)_{cb} = 2 \varepsilon_{abcd}\,.
  \eea
We choose the Minkowski metric $\eta_{MN}$ with a mostly negative
signature. The generators of the spinor representation $\sigma_{MN}$
are real, where
 \bea
 (\sigma_{MN})^a{}_b = \f12 (\tilde{\gamma}^M \gamma^N - \tilde{\gamma}^N \gamma^M)^a{}_b\,.
 \eea

The action of the theory can be derived as a sum of actions for $6D$
vector multiplet and hypermultiplet with specific scalar and Yukawa
interactions and reads\footnote{Some comments about derivation of
this action are given at the end of this section.}
 \bea
 S_{\rm SYM}^{(1,0)} &=&\f{1}{2 {\rm f}^2}\tr\int d^6 x \Big(-F_{MN}F^{MN}
 + i\lambda^{ia}(\gamma^M)_{ab}\nabla_{M}\lambda_{i}^{b}\Big) \nn \\
 && +\f{1}{2 {\rm f}^2}\int d^6 x \Big( -\vp^{\cA i}\,\nabla^2\vp_{\cA i}
 + i \psi^{\cA}_{a}(\tilde{\gamma}^M)^{ab}\nabla_{M}\psi_{\cA b}
 - \f{1}4(\varphi^{\cA i}\varphi_{\cA}^j)^2  - 2\psi^{\cA}_{a}\lambda^{ia}\vp_{{\cA}i} \Big),
 \label{S0}
 \eea
where the hypermultiplet fields are taken in some representation of
the gauge group and $\nabla_M$ is a covariant derivative
$\nabla_M=\partial_M-i A_{M}$. The coupling constant ${\rm f}$ has a
dimension of inverse mass, $[{\rm f}]=-1$. The action \eq{S0} is
invariant under the gauge transformation
 \bea
 \d_{\Lambda} A_M = \nb_M \Lambda\,, &\quad&  \d_{\Lambda} \vp^{\cA}_i = i\Lambda\vp^{\cA}_i\,, \\
 \d_{\Lambda} \lambda^{a}_i = i[\Lambda,\lambda^{a}_i]\,,
 &\quad& \d_{\Lambda} \psi^{\cA}_{a} = i\Lambda\psi^{\cA}_{a}\,,
 \label{gtr}
 \eea
with the gauge transformation parameter $\Lambda = \Lambda(x)$ which
takes value in the Lie algebra of gauge group of the theory in the
vector multiplet and hypermultiplet representations respectively.

The action \eq{S0} possesses $\cN=(1,0)$ supersymmetry. It relates
the spinor field $\psi^{\cA}_{a}$ with the complex scalar
$\vp^{\cA}_i$ and the spinor field $\lambda^{a}_i$ with the gauge
vector field $A_M$ and scalar one $\vp^{\cA}_i$. Explicitly we have
\bea && \delta \vp_{\cA i}=-i\epsilon^{a}_{i}\psi_{\cA a}, \qquad
\qquad
\delta \psi_{\cA a}=\epsilon^{ib}(\gamma^M)_{ba}\nabla_{M}\vp_{\cA i}, \nn \\
&& \delta
A_{M}=\tfrac{i}{2}\epsilon^{ia}(\gamma_M)_{ab}\lambda_{i}^{b},
\qquad \delta
\lambda^{ia}=-\tfrac{1}{2}F^{MN}(\sigma_{MN})^{a}{}_{b}\epsilon^{bi}
+\tfrac{i}{2}\epsilon^{aj}\vp^{\cA i}\vp_{\cA j}, \lb{del4}
\eea
where $\epsilon^{a}_{i}$ is a left-handed parameter of $\cN=(1,0)$
supersymmetry transformation.

If the hypermultiplet component fields $\vp^{\cA i}$ and $ \psi_{\cA
a}$ align in the  adjoint representation of gauge group, the action
\eq{S0} possesses an additional implicite $\cN=(0,1)$ supersymmetry
and describes the six-dimensional $\cN=(1,1)$ SYM theory
 \bea
 S^{(1,1)}_{\rm SYM}=\f{1}{2 {\rm f}^2}\tr\int d^6 x \Big(-F_{MN}F^{MN}
 + i\lambda^{ia}(\gamma^M)_{ab}\nabla_{M}\lambda_{i}^{b}
 - \vp^{\cA i}\,\nabla^2\vp_{\cA i}+\nn \\
 + i \psi^{\cA}_{a}(\tilde{\gamma}^M)^{ab}\nabla_{M}\psi_{\cA b}
 - \f{1}4[\varphi^{\cA i},\varphi_{\cA}^j]^2  - 2\psi^{\cA}_{a}[\lambda^{ia},\vp_{{\cA}i}] \Big).
 \label{S11}
 \eea
Indeed the action \eq{S11} is invariant under both $\cN=(1,0)$
supersymmetry transformations
 \bea
 && \delta \vp_{\cA
i}=-i\epsilon^{a}_{i}\psi_{\cA a}, \qquad \qquad
\delta \psi_{\cA a}=\epsilon^{ib}(\gamma^M)_{ba}\nabla_{M}\vp_{\cA i}, \nn \\
&& \delta
A_{M}=\tfrac{i}{2}\epsilon^{ia}(\gamma_M)_{ab}\lambda_{i}^{b},
\qquad \delta
\lambda^{ia}=-\tfrac{1}{2}F^{MN}(\sigma_{MN})^{a}{}_{b}\epsilon^{bi}
+\tfrac{i}{2}\epsilon^{aj}[\vp^{\cA i},\vp_{\cA j}], \lb{del5} \eea
and $\cN=(0,1)$ ones \bea && \delta_0 A_{M} =
\tfrac{i}{2}\epsilon^{\cA}_a (\tilde{\gamma}_{M})^{ab}\psi_{\cA
b}\,, \qquad \
\delta_0 \vp_{Ai} =-i\epsilon_{Aa} \lambda^{a}_{i}, \nn \\
&& \delta_0 \lambda^{ia} =
\epsilon^{\cA}_{b}(\tilde{\gamma}^{M})^{ba}\nabla_{M}\vp^{i}_{\cA},
\qquad \delta_0 \psi_{\cA a}
=-\tfrac{1}{2}\epsilon_{Ab}(\sigma^{MN})^b{}_a F_{MN}
+\tfrac{i}{2}\epsilon_{\cB a}[\vp^{\cB i},\vp_{\cA i}], \label{del5}
 \eea
mixing the spinor field $\lambda^{ia}$ and the scalar $\vp_{\cA i}$
one and etc. with corresponding right-handed spinor parameter
$\epsilon_{\cA a}$.

The action \eq{S11} of the $\cN=(1,1)$ SYM theory was obtained at
first by dimensional reduction of the corresponding ten-dimensional
SYM theory (see e.g., \cite{MS1},\cite{MS2}). As to the action for
$\cN=(1,0)$ vector multiplet theory interacting with
hypermultiplets, it was known only in superfield form \cite{BIS} and
its complete component form was unknown. It is evident that the
component form of such an action can be useful for various aims. The
action \eq{S0} fills this gap. Let us briefly comment on the
derivation of this action. We follow to Noether procedure beginning
with the free actions for vector multiplet and hypermultiplet and
assuming invariance under the linearized supersymmetry
transformations. Then we include the appropriate interactions and
modify the supersymmetry transformations. As a result, we arrive at
the action \eq{S0} and transformations (\ref{del4}). The action of
$\cN=(1,1)$ theory (\ref{S11}) is obtained from \eq{S0} if to take
the hypermultiplet in the adjoint representation. Then we construct
the transformations (\ref{del5}) and explicitly check that the
action (\ref{S11}) is invariant under these transformations. As far
as we know, the component actions \eq{S0} and \eq{S11} for the
$\cN=(1,0)$ and $\cN=(1,1)$ SYM theories with corresponding
supersymmetry transformation \eq{del4}, \eq{del5} were not obtained
earlier in the framework of component formulation.

\section{Background field method and  one-loop divergences}

We are going to study the effective action in the theory with a
classical action \eq{S0}. To preserve the manifest gauge invariance
we will use the background field method (see e.g., \cite{Weinberg}, \cite{sred}). First of all we split the
initial gauge field $A_M$ and scalar field $\vp_{\cA i}$ to
classical {\it background} fields $\mathbf{A}_{M},
\mathbf{\Phi}_{\cA i}$ and {\it quantum} ones $a_{M}, \vp_{\cA i},
\lambda^{a}_i, \psi_{\cA a} $
\bea
&& A_M \rightarrow {\rm f} a_M + \mathbf{A}_M, \qquad \vp_{\cA i} \rightarrow {\rm f}\vp_{\cA i}
+ \mathbf{\Phi}_{\cA i}, \nn \\
&&\,\,\,\, \lambda^{a}_i \rightarrow  {\rm f} \lambda^{a}_i,\qquad
\qquad \quad \psi_{\cA a} \rightarrow {\rm f}\psi_{\cA a}.
\label{bgr1} \eea We are going to study the bosonic sector of
effective action in the model \eq{S0}. Therefore we introduce the
background field only for $A_M$ and $\vp_{\cA i}$ and keep the
spinor fields $\lambda^{ai}$ and $\psi_{\cA a}$ as a quantum ones.

Following Fadeev-Popov method we choose the gauge fixing action in
the standard way (see e.g., \cite{Weinberg}, \cite{sred})
 \bea
 S_{\rm gf} = - \tr \int
d^6 x (\boldsymbol{\nabla}^M a_M)^2\,. \label{Sgf}
 \eea
Here  $\boldsymbol{\nabla}_M=\partial_{M}-i\mathbf{A}_{M}$ are the
background dependent covariant derivative. The corresponding
Fadeev-Popov ghosts $\bar c$ and $c$ action is
 \bea
 S_{\rm gh} =- \tr \int d^6x\,\, \bar c\,  \boldsymbol{\nabla}^M \Big(\partial_M - i (\mathbf{A} + a)_M\Big)c\,.
 \label{Sgh}
 \eea

The total quantum action $S_{\rm quant}$ constructs as sum of the
classical action \eq{S0} with shifted field variables \eq{bgr1} and
the actions \eq{Sgf} and \eq{Sgh}. According to the background field
method one has to introduce the corresponding background and quantum
gauge transformations for both quantum fields and classical
background ones. These transformations are constructed on the base
of the initial gauge transformation \eq{gtr} and splitting \eq{bgr1}
(see for details, e.g., \cite{Weinberg}, \cite{sred})). The
background field method provides the procedure of constructing the
effective action $\Gamma[{\bf A}, {\bf \Phi}]$ for background fields
${\bf A}_M$ and ${\bf \Phi}_{\cA i}$ which is invariant under the
classical gauge transformations.

\subsection{One-loop divergences in the $\cN =(1,0)$ SYM theory}

Let we proceed to the study of the one-loop divergences in the
$\cN=(1,0)$ SYM theory, where the hypermultiplet transforms under
the arbitrary representation $R$ of the gauge group. We consider the
logarithmic divergences in a sector of gauge field ${\bf A}_M$ and
switch off the background scalar field ${\bf \Phi}_{\cA i}=0$. We
assume the background field ${\bf A}_M$ is completely off-shell. We
choose the gauge group to be compact Lie one and use the Hermitian
basis for generators
 \bea
 [T^{\rm I},T^{\rm J}]=i f^{\rm IJK} T^{\rm K}\,, \qquad \tr(T^{\rm I} T^{\rm J}) = {\rm T_R}\d^{\rm IJ}, \qquad
 \tr(T^{\rm I} T^{\rm J}T^{\rm K}) = \tfrac12 {\rm T_R} f^{\rm IJK} + \tfrac12 {\rm A_R} d^{\rm IJK},
  \eea
where $f^{\rm IJK}$ is a totally antisymmetric structure constants,
$d^{\rm IJK}$ is  a totally symmetric tensor and ${\rm A_R}$ is the
anomaly coefficient of the representation $R$ \cite{sred}. The
generators of the fundamental representation $T_f^{\rm I} \equiv
t^{\rm I}$ are normalized in the standard way $\tr(t^{\rm I} t^{\rm
J}) = \f12 \delta^{\rm IJ}$. In this case the one-loop contribution
to effective action is
 \bea
 \Gamma^{(1)}_{(1,0)}[{\bf A}] &=& \frac{i}{2}\Tr_{\rm Adj}\ln \Big(- \eta_{MN} (\boldsymbol{\nabla}^2)^{\rm IJ}-2f^{\rm IKJ}
 {\bf F}^{\rm K}_{MN}\Big)  -\frac{i}{4}\Tr_{\rm Adj}\ln \Big(   i \d_i{}^j (\gamma^M)_{ab} \boldsymbol{\nabla}_M \Big)^2 \nn \\
 && +  \frac{i}{2}\Tr_{\rm R}\ln \Big( -\d_i{}^j\d_\cA{}^\cB \boldsymbol{\nabla}^2 \Big) -\frac{i}{4}\Tr_{\rm R}\ln \Big( i \d_\cA{}^\cB (\tilde\gamma^M)^{ab} \boldsymbol{\nabla}_M\Big)^2
\nn \\
&& -i\Tr_{\rm Adj}\ln \Big(-\boldsymbol{\nabla}^2\Big),
 \label{Gamma2}
 \eea
where ${\bf F}^{\rm K}_{MN}$ is a background vector field strength
and we have denoted $\boldsymbol{\nabla}^2 = \boldsymbol{\nabla}^M
\boldsymbol{\nabla}_M$. The first two contributions in \eq{Gamma2}
come from vector multiplet. The next two terms in the second line of
\eq{Gamma2} are the contributions from hypermultiplet which
transforms under some irreducible representation $R$ of the gauge
group $G$. The last term in \eq{Gamma2} is a contribution from the
ghosts fields.

We use proper time method to study one-loop effective action. For
differential operators $\Delta$ associated with action \eq{Gamma2} we
define
 \bea
 \f{i}2\Tr \ln \Delta = -\f{i}2\mu^{2\omega} \tr \int d^6 x \int_0^\infty \f{d (is)}{(is)^{1-\omega}}
 e^{is \Delta} \delta^6(x-x')\Big|_{x=x'},
 \label{proper}
 \eea
where $\mu$ is an arbitrary parameter of mass dimension and trace is
taken over all gauge group, spinor and Minkowski space-time indexes.
Also we introduce  the parameter of dimension regularization
$\omega$. Logarithmic divergent contributions in \eq{Gamma2} arise
as a pole $\frac{1}{\omega}$ in the limit $\omega \to 0$. To
calculate such terms we use \eq{proper} and expand the exponent of
each operator up to sixth order over covariant derivative
$\boldsymbol{\nabla}_M$. After that we pass to momentum
representation for delta-function and calculate the integral over
momentum and proper time $s$. The result for each contribution in
\eq{Gamma2} is listed below
 \bea
 \frac{i}{2}\Tr\ln \Big(-\eta_{MN} (\boldsymbol{\nabla}^2)^{\rm IJ}-2f^{\rm IKJ} {\bf F}^{\rm K}_{MN}\Big)\Big|_{\rm div}
 &=& -\frac{C_2}{(4\pi)^3 \omega}\Big(\frac{3}{90}F_3 + \frac{17}{60}I_3\Big),
 \label{div1} \\
 \frac{i}{2}\Tr\ln \Big( -\d_i{}^j\d_\cA{}^\cB \boldsymbol{\nabla}^2\Big)\Big|_{\rm div}
 &=& - \frac{\rm T_R}{(4\pi)^3 \omega} \Big(\frac{2 }{90}F_3-\frac{ 2 }{60}I_3\Big)
 - \frac{\rm A_R}{(4\pi)^3 \omega} \frac{2}{90}\bar F_3, \label{3.9}\\
  -\frac{i}{4}\Tr\ln \Big(i \d_i{}^j (\gamma^M)_{ab} (\boldsymbol{\nabla}_M\Big)^2\Big|_{\rm div}
  &=& \frac{C_2}{(4\pi)^3 \omega} \Big(\frac{2}{90}F_3+\frac{2}{15}I_3\Big),\\
 -\frac{i}{4}\Tr\ln \Big(i \d_\cA{}^\cB  (\tilde\gamma^M)^{ab} (\boldsymbol{\nabla}_M)^{\rm IJ}\Big)^2\Big|_{\rm div}
  &=& \frac{\rm T_R}{(4\pi)^3 \omega} \Big(\frac{2}{90}F_3+\frac{2}{15}I_3\Big)
  +\frac{\rm A_R}{(4\pi)^3 \omega} \frac{2}{90}\bar F_3, \label{3.11}\\
 -i\Tr\ln\Big(-(\boldsymbol{\nabla}^2)^{\rm IJ}\Big)\Big|_{\rm div}
  &=& \frac{C_2}{(4\pi)^3 \omega}\, \Big(\frac{1}{90}F_3-\f{1}{60}I_3\Big), \label{div2}
 \eea
where $C_2$ is a second Casimir operator of adjoint representation
of gauge group and we have introduced the quantities
 \bea
 && F_3 = \int d^6x f^{\rm IJK}\big({\bf F}_M{}^N\big)^{\rm I}
 \big({\bf F}_N{}^L\big)^{\rm J}\big({\bf F}_L{}^M\big)^{\rm K}, \qquad
 I_3 = \int d^6x \big(\boldsymbol{\nabla}_M {\bf F}^{M L} \big)^{\rm I}
 \big(\boldsymbol{\nabla}^N {\bf F}_{N L}\big)^{\rm I}\,, \nn \\
 && \bar F_3 = \int d^6x d^{\rm IJK}\big({\bf F}_M{}^N\big)
 ^{\rm I}\big({\bf F}_N{}^L\big)^{\rm J}\big({\bf F}_L{}^M\big)^{\rm K}.
 \label{F3}
 \eea
Summing up all divergent contributions  \eq{div1} - \eq{div2} to the
one-loop effective action  \eq{Gamma2} one can see that all
contributions with $F_3$ and $\bar F_3$ cancel each other and the
divergent part of effective action is proportional to the equation
of motion for background vector field
 \bea
 \Gamma^{(1)}_{\rm div}[{\bf A}] = \f{{\rm T_R}-C_2}{3 (4\pi)^3 \, \omega}
 \tr \int d^6x \boldsymbol{\nabla}_M {\bf F}^{M L} \boldsymbol{\nabla}^N {\bf F}_{N L}. \label{Gamma2_1}
 \eea
The absence of divergent contributions with $F_3$ and $\bar F_3$ in
the $\cN=(1,0)$ SYM theory without hypermultiplet was mentioned
earlier \cite{HRT}.  Here we demonstrate absence of these terms by
explicit calculation of the divergences in theory including the
hypermultiplet in arbitrary representation of the gauge group.  The
divergent contribution from quantum hypermultiplet \eq{3.9} and
\eq{3.11} containing $F_3$ and $\bar F_3$ exactly cancel each other.
Also we note that this result is consistent with the superfield
calculations \cite{BIMS2,BIMS3}. The divergent part of one-loop
superfield effective action in gauge multiplet sector includes the
square of classical superfield equations of motion and the
contributions with $(F_{MN})^3$ in the components are ruled out.

In the $\cN=(1,1)$ SYM theory the hypermultiplet transforms under
adjoin representation of gauge group. In this case ${\rm T_R} = C_2$
and ${\rm A_R} = 0$ and we obtain the cancelation of ultraviolet
divergences in \eq{Gamma2_1} off-shell \cite{Kazakov,BIMS2,BIMS3}.
However result is valid only for Fermi-Feynman gauge and in case of
general off-shell background it can be gauge-dependent
\cite{Kazakov,BIMS5}. It means that in general the one-loop
divergences of the theory under consideration apparently are absent
only on-shell.

\subsection{One-loop effective action in $\cN=(1,1)$ SYM theory}

In this subsection we consider the finite one-loop contribution to the
effective action for the $\cN=(1,1)$ SYM theory. Unlike the previous section,
the background scalar field is also taken into account.

One-loop
contribution $\Gamma^{(1)}[{\bf A}, {\bf \Phi}]$ is determined by
the quadratic over quantum fields part of quantum action
 \bea
 S^{(1,1)}_2[{\bf A},{\bf \Phi}] &=& \f12\tr \int d^6 x \Big(
   2 a^M \boldsymbol{\nabla}^2 a_M
   + 4i a^M[\mathbf{F}_{MN},a^N]
   -[a^M,\mathbf{\Phi}^{\cA i}]^2
   - 2i [a^M, {\bf \Phi}^{\cA i}]\boldsymbol{\nabla}_M \vp_{\cA i}
   \nn \\
&&
 -\vp^{\cA i}\boldsymbol{\nabla}^2 \vp_{\cA i}
 - [\vp^{\cA i},{\bf \Phi}_{\cA}^j]^2 -\f{1}2[\vp^{\cA i},\vp_{\cA}^j]
 [{\bf \Phi}^{\cB}_i,{\bf \Phi}_{\cB j}]
  -2i[a^M, \vp^{\cA i}]  \boldsymbol{\nabla}_M {\bf \Phi}_{\cA i}   \nn \\
&& + i\lambda^{i a} (\gamma^M)_{ab}\boldsymbol{\nabla}_M
\lambda^{a}_i
 + i\psi^{\cA}_a(\tilde{\gamma}^M)^{ab} \boldsymbol{\nabla}_M \psi_{\cA b}
 - 2\psi^\cA_a[\lambda^{ia},\mathbf{\Phi}_{\cA i}] - 2 \bar c \boldsymbol{\nabla}^2 c \Big).
    \label{S2}
 \eea
We  note that the quadratic action $\eq{S2}$ contains terms
mixing quantum vector $a_M$ and scalar $\vp_{\cA i}$ fields as well as the
quantum spinor
$\lambda^{ia}$ and $\psi^\cA_a$ fields. In order to exclude the mixed $a_M$
and $\vp_{\cA i}$ terms we introduce the gauge fixing action in
form of the $R_{\xi}$ gauge (see e.g., \cite{Weinberg}, \cite{sred})
 \bea
 S_{\rm gf} = - \tr \int d^6x (\boldsymbol{\nabla}_M a^M - \tfrac{i}2 [\mathbf{\Phi}^{\cA i}, \varphi_{\cA i}])^2\,.
 \label{Sgf2}
 \eea
Using the gauge fixing action \eq{Sgf2} instead of \eq{Sgf} leads to
cancelation of mixed $a_M$ and $\vp_{\cA i}$ terms in quadratic
action \eq{S2}. However in that case the additional contribution
arises for the ghosts fields.

Integrating over quantum fields in the functional integral we
produce the one-loop contribution to the effective action
 \bea
 \Gamma^{(1)}_{(1,1)}[{\bf A}, {\bf \Phi}]
 =\frac{i}{2}\Tr\ln \cB-\frac{i}{4}\Tr\ln\mathbf \cF^2
 -i\Tr\ln{\cal G},
 \label{Gamma1}
 \eea
where ${\cal G} = -(\boldsymbol{\nabla}^2)^{\rm IJ} + \tfrac12
f^{\rm IKP}f^{\rm JLP} ({\bf \Phi}^{\cA i})^{\rm K}({\bf \Phi}_{\cA
i})^{\rm L}$. The matrices $\cB$ and $\cF$ depend on the background
fields $\boldsymbol{A}_M$ and $\boldsymbol{\Phi}_{\cA i}$. To reveal
the explicit structure of the matrices  $\cB$ and $\cF$ it is useful
to write down the the gauge group indices. All
fields we assumed in adjoint representation of gauge group.  We
represent the matrices $\cB$ and $\cF$ in a block form \bea
\cB=\begin{pmatrix}
\cB_1 & 0 \\
0 & \cB_2
\end{pmatrix}, \qquad
\cF=\begin{pmatrix}
\cF_1 & \cF_2 \\
\cF_3 & \cF_4
\end{pmatrix},
\eea where for the $\cB$ matrix elements we have introduced the
notations
 \bea
 (\cB_1^{\rm IJ})_{MN} &=& -\eta_{MN}
 (\boldsymbol{\nabla}^2)^{\rm IJ}-2f^{\rm IKJ} {\bf F}^{\rm K}_{MN}
 +\tfrac12\eta_{MN} f^{\rm IKP}f^{\rm JLP}
  ({\bf \Phi}^{\cA i})^{\rm K}({\bf \Phi}_{\cA i})^{\rm L}, \\
  (\cB_2^{\rm IJ})_{\cA i}{}^{\cB j} &=& - \tfrac12\d_i{}^j\d_\cA{}^\cB
 (\boldsymbol{\nabla}^2)^{\rm IJ}
-\tfrac{1}2 \d_i{}^j f^{\rm IKP}f^{\rm JLP}({\bf \Phi}^k_\cA)^{\rm
K}
({\bf \Phi}^{\cB}_k)^{\rm L} \nn \\
&& -\tfrac{1}4 \d_\cA{}^\cB f^{\rm IJP}f^{\rm PKL}({\bf \Phi}^{\cal
C}_i)^{\rm K} ({\bf \Phi}_{\cal C}^j)^{\rm L}\,,
 \eea
and for the $\cF$ elements  we have the following expressions
 \bea
 (\cF_1^{\rm IJ})_{ab}{}_i^j &=& \tfrac{i}2 \d_i{}^j (\gamma^M)_{ab}
 (\boldsymbol{\nabla}_M)^{\rm IJ}, \\
 (\cF_2^{\rm IJ})_a^b{}^\cB_i &=& -\tfrac{i}{2}\d_a{}^b f^{\rm IKJ}
 ({\bf \Phi}^{\cB}_i)^{\rm K},\\
 (\cF_3^{\rm IJ})^a_b{}_\cA^j &=& -\tfrac{i}{2}\d^a{}_b f^{\rm IKJ}
 ({\bf \Phi}_{\cA}^j)^{\rm K},\\
 (\cF_4^{\rm IJ})^{ab}{}_\cA^\cB &=& \tfrac{i}2 \d_\cA{}^\cB
 (\tilde\gamma^M)^{ab} (\boldsymbol{\nabla}_M)^{\rm IJ}.
 \eea
Equations of motion for background fields has the following form
 \bea
 \boldsymbol{\nabla}^M F_{MN} + \tfrac{i}2[{\bf \Phi}^{\cA i},
 \boldsymbol{\nabla}_N {\bf \Phi}_{\cA i}] &=& 0, \nn \\
 \boldsymbol{\nabla}^2 {\bf \Phi}^{\cA i} -\tfrac{ 1}{2} {\bf \Phi}^{\cA}_j
 [{\bf \Phi}^{\cB i}, {\bf \Phi}^{j}_\cB] &=& 0. \label{EQM1}
 \eea

The one-loop divergences in the theory under consideration were
studied in many details both in component and superfield approaches
\cite{MS1,MS2,FT,HS,HS1,BHS,BHS1,Kazakov,BIMS2,BIMS3,HRT}. The
absence of  one-loop divergences  in the six-dimensional $\cN=(1,1)$
SYM theory on-shell was known \cite{MS1,MS2,FT} and, as we discussed
in the previous section, it is immediately following from the
\eq{Gamma2_1}.

\section{Leading low-energy contribution to one-loop
effective action of $\cN=(1,1)$ SYM theory}

Next we proceed to the evaluation of leading low-energy finite
contribution to effective action \eq{Gamma1}. In what follows we
assume the gauge group of the theory to be $SU(2)$. We choose
standard Hermitian basis for generators $\tau^{\rm I} =\f12
\sigma^{\rm I}$ using Pauli matrices
 \bea
 \sigma^{1} =\begin{pmatrix}
0& 1 \\
1& 0
\end{pmatrix}, \quad
 \sigma^{2}  =\begin{pmatrix}
0& -i \\
i& 0
\end{pmatrix}, \quad
 \sigma^{3}  =\begin{pmatrix}
1& 0\\
0& -1
\end{pmatrix}. \quad
 \eea
Generators $\tau^{\rm I}, \, {\rm I}=1,2,3,$ satisfy the $su(2)$
algebra, $[\tau^{\rm I}, \tau^{\rm J}] = i \epsilon ^{\rm IJK}
\tau^{\rm K}$, with totally antisymmetric symbol $\epsilon ^{\rm
IJK}$ and they are normalized as, $\tr (\tau^{\rm I} \tau^{\rm J}) =
\f12 \d^{\rm IJ}$.

We also suppose that background fields ${\bf A}_M$ and ${\bf
\Phi}_{\cA i}$ align into Cartan subalgebra of $su(2)$ generated by
$\tau^3$ matrix
 \bea
 {\bf A}_M = A_M \tau^3, \qquad {\bf\Phi}_{\cA i} = \Phi_{\cA i} \tau^3.
 \label{bg1}
 \eea
We have denoted by $A_M$ and $\Phi_{\cA i}$ the third components of
background fields ${\bf A}_M$ and ${\bf \Phi}_{\cA i}$. Our choice
of background corresponds to the case when gauge symmetry group
$SU(2)$ is broken to $U(1)$. Equations of motion for background
fields \eq{EQM1} in accordance with restriction \eq{bg1} take a free
form
 \bea
 \partial^M F_{MN}=0, \qquad \partial^2 \Phi_{\cA i} = 0,
 \label{EQM2}
 \eea
where $F_{MN} = \partial_M A_N - \partial_N A_M$ is an Abelian field
strength corresponding to ${\bf F}_{MN} = F_{MN}\tau^3$. Also we
assume that background fields are slowly varying in space-time
 \bea
 \partial_M F_{NL} = 0, \qquad \partial_M \Phi_{\cA i} = 0.
 \label{bg2}
 \eea
The last condition means that we systematically neglect all terms
with derivatives of the background field strength $F_{MN}$ and the
scalar one $\Phi_{\cA i}$.

Gauge transformations for background fields in case of \eq{bg1} are
simplified and take the form
 \bea
\d_\Lambda A_M = \partial_M \Lambda, \qquad \d_\Lambda \Phi_{\cA i}
= 0, \label{BGgauge2}
 \eea
where $\Lambda = \Lambda(x)$ is an Abelian gauge parameter. In
Abelian case the gauge field strength $F_{MN}$ is known to be
invariant under gauge transformation \eq{BGgauge2}, $\d_\Lambda
F_{MN}=0$. Hence the effective action as a polynomial function of
$F_{MN}$ is a gauge invariant under construction.

\subsection{Leading low-energy contribution}

First of all let we rewrite the matrices $\cB$ and $\cF$ in
accordance with our choice of background fields \eq{bg1} and
\eq{bg2}. We consider non-vanishing background for scalar field
$\Phi_{\cA i}$, hence we have
  \bea
 \Gamma^{(1)}[A, \Phi]
 &=&\frac{i}{2}\Tr\ln \Big( -\eta_{MN}(\nb^2)^{\rm IJ}
 +\tfrac12\eta_{MN} \d^{\rm IJ} \Phi^2
 - 2\epsilon^{\rm I3J} F_{MN} \Big)\nn \\
 && + \frac{i}{2}\Tr\ln \Big(-\d_i{}^j\d_\cA{}^\cB (\nabla^2)^{\rm IJ}
+\tfrac12\d_i{}^j \d_\cA{}^\cB \d^{\rm IJ}\Phi^2\Big)  \nn \\
&&-\frac{i}{4}\Tr\ln \Big(-\d_i{}^j (\nb^2)^{\rm
IJ}-\tfrac{i}{2}\d_i{}^j\epsilon^{\rm I3J}\sigma^{MN}F_{MN}
+\tfrac{1}{2}\d_i{}^j\d^{\rm IJ} \Phi^2\Big) \nn \\
&&-\frac{i}{4}\Tr\ln \Big(-\d_\cA{}^\cB(\nb^2)^{\rm
IJ}-\tfrac{i}{2}\d_\cA{}^\cB\epsilon^{\rm I3J}\sigma^{MN}F_{MN}
+ \tfrac{1}{2} \d_\cA{}^\cB\d^{\rm IJ} \Phi^2 \Big) \nn \\
 && -i\Tr\ln\Big(-(\nb^2)^{\rm IJ}+\tfrac12\d^{\rm IJ} \Phi^2\Big),
 \label{Gamma3}
 \eea
where the operator $(\nb^2)^{\rm IJ}$ is constructed using the
Abelian  covariant derivative $(\nb_M)^{\rm IJ} = \d^{\rm
IJ}\partial_M + \epsilon^{\rm I3J} A_M$ and we introduced $\Phi^2 =
\Phi^{\cA i} \Phi_{\cA i}$. In order to obtain the leading
low-energy contribution to one-loop effective action \eq{Gamma3} one
should follow the same method as in study of divergences in previous
section. We use the proper time technique $\eq{proper}$ to study the
one-loop effective action \eq{Gamma3}. We are interested in the
leading low-energy contributions to one-loop effective action which
have not higher than fourth power of gauge field strength $F_{MN}$.
Thus we have to expand the exponent in the $\eq{proper}$ and collect
all terms up to the eighth power of covariant derivative or the
forth power of background gauge field strength. In order to
determine the structure of the leading contribution to effective
action we consider the contribution of forth power over background
gauge field strength from the determinant of spinor field
$\lambda^{i a}$
 \bea
-\frac{i}{4}\Tr\ln \Delta_{1/2}  = \frac{i}{4}\mu^{2\omega}\tr \int
d^6 x \int_0^\infty \f{d (is)}{(is)^{1-\omega}} e^{is \Delta_{1/2}}
\delta^6(x-x')\Big|_{x=x'},
 \eea
where $\Delta_{1/2} = -\d_i{}^j (\nb^2)^{\rm
IJ}-\tfrac{i}{2}\d_i{}^j\epsilon^{\rm I3J}\sigma^{MN}F_{MN}
+\tfrac{1}{2}\d_i{}^j\d^{\rm IJ} \Phi^2$. We assume the constant
background scalar field, thus the covariant d'Alembertian in the
exponent is comute with the $\Phi^2$ term
 \bea
-\frac{i}{4} \Tr\ln \Delta_{1/2}  &=& \frac{i}{4}\tr \int d^6 x \int_0^\infty \f{d (is)}{(is)} e^{\tfrac{i}{2}s \Phi^2}e^{-is\nb^2 -\tfrac{i}{2}s \sigma_{MN} F^{MN}} \delta^6(x-x')\Big|_{x=x'} \nn \\
&=& \f{i}{24} \int d^6 x \, \tr \big(\tfrac12  \sigma_{MN} F^{MN}
\big)^4 \int_0^\infty d (is)\,(is)^3 e^{\tfrac{i}{2}s \Phi^2}
e^{-is\nb^2}\delta^6(x-x')\Big|_{x=x'}+...,\nn
 \eea
where the trace is taken over spinor indexes and dots mean over
contributions of the forth power of $F_{MN}$. After that we pass to
momentum space for delta-function and calculate the integral over
proper-time. We have
 \bea
-\frac{i}{4} \Tr\ln \Delta_{1/2} \sim \frac{1}{(4\pi)^3}
 \int d^6 x\frac{1}{\Phi^2}\Big(4F^4-3F^2 F^2\Big),
 \eea
where we  denote  $F^4 = F_{M}{}^N F_N{}^P F_P{}^Q F_Q{}^M$ and $F^2
= F_{M}{}^N F_N{}^M$. The same strategy we apply to the remaining
terms in \eq{Gamma3}.  We omit tedious details of calculations and
the result is
 \bea
 \Gamma^{(1)}_{\rm lead}[A, \Phi]= \frac{1}{(4\pi)^3}\frac{1}{360}
 \int d^6 x\frac{1}{\Phi^2}\Big(4F^4-3F^2 F^2\Big).
 \label{lead}
 \eea
Recently \cite{BIM} the leading low-energy contribution to the
one-loop effective action in $\cN=(1,1)$ SYM theory was analyzed
using superfield approach. The effective action \eq{lead} coincides
with bosonic sector the superfield effective action obtained in
\cite{BIM}. The component calculation, carried out here, can be
considered as an independent test of the superfield result.

\subsection{Effective potential}
In order to evaluate the one-loop contribution to the effective
action for scalar field $\Phi_{\cA i}$ we switch off the vector
background field $A_M = 0$ in the effective action  \eq{Gamma3}.
Then we obtain
 \bea
 V^{(1)}[\Phi]
 &=&\frac{i}{2}\Tr\ln \Big( -\eta_{MN}\d^{\rm IJ} \partial^2
 +\tfrac12\eta_{MN} \d^{\rm IJ} \Phi^2 \Big)\nn \\
 && + \frac{i}{2}\Tr\ln \Big(-\d_i{}^j\d_\cA{}^\cB \d^{\rm IJ} \partial^2
+\tfrac12\d_i{}^j \d_\cA{}^\cB \d^{\rm IJ}\Phi^2\Big)  \nn \\
&&-\frac{i}{4}\Tr\ln \Big(-\d_i{}^j\d^{\rm IJ} \d_a{}^b\partial^2
+\tfrac12\d_i{}^j\d^{\rm IJ}\d_a{}^b \Phi^2\Big) \nn \\
&&-\frac{i}{4}\Tr\ln \Big(-\d_\cA{}^\cB\d^{\rm IJ}\d^a{}_b
\partial^2 + \tfrac12 \d_\cA{}^\cB\d^{\rm IJ} \d^a{}_b\Phi^2 \Big) \nn \\
&& -i\Tr\ln\Big(-\delta^{\rm IJ}\partial^2+\tfrac12\d^{\rm IJ}
\Phi^2\Big).
 \label{Gamma4}
 \eea
Note that indices of adjoin representation ${\rm I,J} = 1,2$ in the
expression \eq{Gamma4}.

The direct evaluation of the above functional determinants on the
base of proper-time technique leads to cancelation of all
contributions in effective potential \eq{Gamma2}
 \bea
  V^{(1)}[\Phi] =0\,.
 \eea
The absence of the one-loop effective potential for scalar fields is
in consistence with the $\cN=(1,1)$ supersymmetry. The same result
can be explained explicitly using $\cN=(1,0)$ harmonic superfield
consideration in the $\cN=(1,1)$ SYM theory \cite{BIM}. In terms of
superfields the  effective potential for background scalar
superfields corresponds to the contributions which depend only on
constant background hypermultiplet. But all such contributions are
forbidden by the on-shell hidden $\cN=(0,1)$ supersymmetry
\cite{BIM}. The absence of the one-loop contribution to the
effective potential for scalar field in $\cN=(1,1)$ SYM theory is
also dictated by the absence of the Coulomb branch in the theory
under consideration \cite{SW}. However, as far as we know it was
never confirmed before by the direct calculation.

\section{Summary}

In this paper we have studied the bosonic sector of the low-energy
effective action in the six-dimensional $\cN=(1,0)$ and $\cN=(1,1)$
SYM theories \eq{S0} and \eq{S11} formulated in terms of physical
components. First of all we have derived the component action of
$6D,\, \cN=(1,0)$ SYM theory and the corresponding supersymmetry
transformations. The one-loop effective action is considered on the
base the background field method with help of $R_{\xi}$-type gauge.
We provided the explicit calculation of the divergent contributions
to the one-loop effective action in $\cN=(1,0)$ SYM theory
interacting with hypermultiplets. We have demonstrated by direct
calculation of corresponding contributions to the one-loop effective
action that all $F^3$-type divergent contributions (\ref{F3}) cancel
each other. The result is in an agreement with the earlier component
calculations \cite{MS1,MS2,FT,HS,HS1,BHS,BHS1,Kazakov,HRT} without
hypermultiplet contributions and with the superfield analysis
\cite{BIMS2,BIMS3,BIMS4}. We also demonstrated that divergent
contributions in the $\cN=(1,1)$ theory cancel each other and the
theory is one-loop off-shell finite as expected.

Studying the finite contributions to the $\cN=(1,1)$ SYM theory we
have considered the gauge group $SU(2)$ and have assumed that the
background fields take the value in Cartan subalgebra of $su(2)$. To
construct the finite low-energy contribution to the effective action
of the theory we assume the non-zero values for both vector and
scalar background fields and using the $R_{\xi}$ gauge. In case of
slowly varying on-shell background fields we calculated the leading
low-energy contribution to the one-loop effective action. It is
determined by the contributions of the fourth power of the gauge
field strength $F_{MN}$. The obtaining result being in agreement
with the bosonic part of the leading contribution to the one-loop
harmonic superfield effective action \cite{BIM} and it can be
considered as an independent test of the superfield calculation. We
also demonstrate the absence of the one-loop contribution to the
effective potential for the background scalar field, which as we
know was not mentioned before by direct calculations.

As a further development of the approach under consideration we plan
to study the divergences in the fermionic sector and investigate the
two-loop counterterms in the theory \eq{S0}. Also it would be
interesting to study the finite low-energy contributions to the
effective action in high derivative SYM theory \cite{ISZ,CT}.

\subsection*{Acknowledgements}
The authors acknowledge a partial support from the RFBR grant,
project No 18-02-00153. Their research was also supported in parts
by Russian Ministry of Science and High Education, project No.
3.1386.2017.  The work of A.S.B. and B.S.M. was supported in part by
the Russian Federation President grant, the project MK-1649.2019.2.

\end{document}